 \documentclass[final,authoryear,1p,times]{elsarticle}

\usepackage{graphicx}

\usepackage{amssymb}

\usepackage{color}   
\usepackage{amsmath}

\journal{Advances in Space Research}

\DeclareMathOperator{\const}{const}

\begin{document}

\begin{frontmatter}



\title{\textbf{Doubts about the crucial role of the rising-tube mechanism in the
formation of sunspot groups}}


\author[avg]{A. V. Getling\corref{cor}}
\address[avg]{Skobeltsyn Institute of Nuclear Physics, Lomonosov Moscow State
University, Moscow, 119991 Russia} \cortext[cor]{Corresponding
author} \ead{A.Getling@mail.ru}

\author[ri]{R. Ishikawa}
\address[ri]{Hinode Science Center, National Astronomical Observatory of
Japan, 2-21-1 Osawa, Mitaka, Tokyo, 181-8588 Japan}
\ead{ryoko.ishikawa@nao.ac.jp}

\author[aab]{A. A. Buchnev}
\address[aab]{Institute of Computational Mathematics and Mathematical Geophysics,
Novosibirsk, 630090 Russia} \ead{baa@ooi.sscc.ru}

\begin{abstract}
Some preliminary processing results are presented for a dataset
obtained with the Solar Optical Telescope on the \emph{Hinode}
satellite. The idea of the project is to record, nearly
simultaneously, the full velocity and magnetic-field vectors in
growing active regions and sunspot groups at a photospheric
level. Our ultimate aim is to elaborate observational criteria
to distinguish between the manifestations of two mechanisms of
sunspot-group formation --- the rising of an $\Omega$-shaped
flux tube of a strong magnetic field and the in situ
amplification and structuring of magnetic field by convection
(the convective mechanism is briefly described).

Observations of a young bipolar subregion developing within AR
11313 were carried out on 9--10 October 2011. During each 2-h
observational session, 5576-\AA\ filtergrams and Dopplergrams
were obtained at a time cadence of 2 min, and one or two
32-min-long spectropolarimetric fast-mode scans were done.
Based on the series of filtergrams, the trajectories of corks
are computed, using a technique similar to but more reliable
than local correlation tracking (LCT), and compared with the
magnetic maps. At this stage of the investigation, only the
vertical magnetic field and the horizontal flows are used for a
qualitative analysis.

According to our preliminary findings, the velocity pattern in
the growing active region has nothing to do with a spreading
flow on the scale of the entire bipolar region, which could be
expected if a tube of strong magnetic field emerged. No violent
spreading flows on the scale of the entire growing magnetic
region can be identified. Instead, normal mesogranular and
supergranular flows are preserved. Signs of small-scale
structuring of the magnetic field can be detected in the area
where new spots develop, and signs of the presence of
separatrices between the magnetic polarities can be found, such
that the surface flows converge to but not diverge from these
separatrix curves. The observed scenario of evolution seems to
agree with Bumba's inference that the development of an active
region does not entail the destruction of the existing
convective-velocity field. The convective mechanism appears to
be better compatible with observations than the rising-tube
mechanism.

In the umbras of the well-developed sunspots, flows converging
to the umbra centres are revealed. Spreading streams are
present around these spots.
\end{abstract}


\end{frontmatter}


\section{Introduction}
\label{intr}

The idea that the magnetic field of a bipolar sunspot group
originates from the emergence of an $\Omega$-shaped flux-tube
loop was considered to be virtually indisputable for a long
time. This \emph{rising-tube model} assumes that the flux tube,
formed in the general solar toroidal magnetic field deep in the
convection zone, carries upward the field that is assumed to be
originally strong. This mechanism received much attention after
the well-known study by \citet{parker}, who invoked magnetic
buoyancy to account for the rise of flux-tube loops, and has
been revisited by a number of investigators over several
decades. Interesting analyses of this model were made, in
particular, by \citet{clgretal1,clgretal2}; numerical
simulations of this mechanism based on full systems of MHD
equations have also been carried out \citep[see, e.g.,][and
references therein]{fanetal,rempcheung}.

\begin{figure} 
\centering
\includegraphics[bb=0 0 504 285pt,clip,
width=0.8\textwidth]{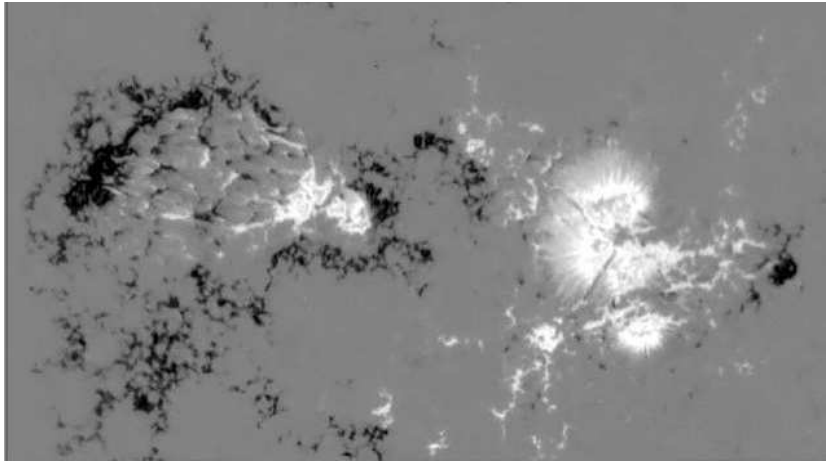}
\caption{A magnetogram of the ``trilobite'' series
demonstrating the emergence of sunspot 10926. This series was
obtained with the Solar Optical Telescope
on the \emph{Hinode} satellite in December 2006. Black
represents negative polarity, and white represents positive.
Adapted from the movie available via a link in \citet{nasa}.\protect}
\label{tril}
\end{figure}

In recent years, however, the rising-tube model has more and
more been objected based on both recent, very detailed
observational data and current views of the processes in the
deep layers. First of all, if it is adopted, one has to account
for the origin of the coherent tube of strong magnetic field;
to this end, some additional, fairly artificial assumptions
need to be introduced. Second, a very important point is that
the pattern inferred from this model for photospheric flows and
magnetic fields on the scale of the growing magnetic region
disagrees with the pattern actually observed on the Sun
\citep[see, e.g.,][and the brief description of a remarkable
magnetogram given below in this Introduction]{kosov}; we shall
see that the results of the present study also scarcely conform
to the scenario that should be expected in the case of
emergence of a large tube of strong field. It was also noted
that the behaviour of the tilt angle of the bipolar magnetic
regions is not consistent with the predictions based on the
rising-tube model \citep{kosovstenflo,kosov}.

We postpone summarising the weak points of the rising-tube
model to the conclusive section of our paper and give now only
an impressive example of the observed features that do not
support this model. This example can be found in the movie
showing the so-called solar-``trilobite'' magnetogram series,
which was obtained on the \emph{Hinode} satellite using the
Solar Optical Telescope (SOT) in nearly the same way as the
magnetograms considered here \citep[see][ and
Section~\ref{obs}]{kosugietal,tsunetaetal}, in December 2006
\citep{nasa}. As noted in the discussion at a ``Living with a
Star'' workshop (September 2007; see the cited webpage of NASA
news), these \emph{Hinode} observations of emergent sunspot
10926 ``challenge traditional views of sunspot formation.''
Specifically, the trilobite data show a sunspot-formation
process differing from what occurs if ``a `rope' of strong
magnetic field beaches the visible surface of the Sun.''
According to Marc DeRosa (Lockheed Martin Solar and
Astrophysics Laboratory in Palo Alto, Calif.), who coined the
name \emph{trilobite} for the feature in question, ``The
emergence of the sunspot magnetism progressed in a very complex
manner, with small pieces appearing to self-assemble into
larger, more coherent structures.''

Indeed, our inspection of the ``trilobite'' movie shows that,
although spreading flows related to the emerging magnetic field
can be noted in this movie, they are finely structured and form
cells resembling convection cells rather than a unique flow
system on the scale of the entire magnetic region (see the
upper left quadrant of Fig.~\ref{tril}). Furthermore, spreading
flows are associated locally with each developing magnetic
island rather than ``globally'' with the entire complex
magnetic configuration (which could be expected if a tube
emerged). In its appearance, such spreading is similar to the
flow around an effervescent tablet on the water surface. An
example of such a ``tablet'' is the magnetic feature in the
right-hand side of the magnetogram, which resembles the
trilobite fossil animal.\footnote{Strictly speaking, the
feature resembling a trilobite emerged somewhat later than the
magnetogram shown in Fig.~\ref{tril} was taken. We use the word
``trilobite'' simply as an identifier of the whole series of
images under discussion.}

Thus, the rising-tube mechanism no longer appears to determine
a paradigm in the investigation of the development of local
magnetic fields. On the other hand, alternative mechanisms have
been suggested in the literature.

Before mentioning some of them, let us introduce necessary
terminological conventions. While the rising-tube mechanism
assumes the original presence of an intense flux tube of the
global toroidal magnetic field in deep layers, other known
mechanisms do not require any strong initial field ensuring
some in situ magnetic-field amplification and structuring. We
restrict ourselves only to processes on scales no larger than
the size of an active region; they will be referred to here as
\emph{local mechanisms}. On the other hand, local processes on
scales small compared with the granular size, which do not play
any decisive role in the formation of active regions, will also
fall beyond our scope of attention. Local MHD mechanisms with
inductive self-excitation of magnetic fields strongly coupled
with fluid motions are of particular interest, since there is a
variety of possibilities for their manifestation under solar
conditions. They are usually termed \emph{local dynamo
mechanisms}. If the local dynamo action is due to thermal
convection, we shall assign such a dynamo to \emph{local
convective dynamos}; they will be discussed in the next
section.

The idea of local MHD dynamo traces back to \citet{gurleb}, who
related the amplification process to the effects of plasma
motions; however, they did not attributed these motions to
convection and even did not specify any particular type of
motion.

\citet{kitmaz} investigated a hydromagnetic instability that
can act on scales large compared to the granular size,
producing a magnetic-flux concentration similar to those
observed in sunspots. The process crucially depends on the
presence of fluid motion and on the quenching of eddy
diffusivity by the enhanced magnetic field with the plasma
cooling down. This is a local mechanism, which, however, is not
a dynamo in the strict meaning of this term.

Another local mechanism, which also does not qualify as a
dynamo, is related to the so-called
negative-effective-magnetic-pressure instability (NEMPI), see
\citet{warnetal} and references therein. It results from the
suppression of the total turbulent pressure (the sum of
hydrodynamic and magnetic components) by the magnetic field.

\begin{figure} 
\centering
\includegraphics[width=0.6\textwidth]{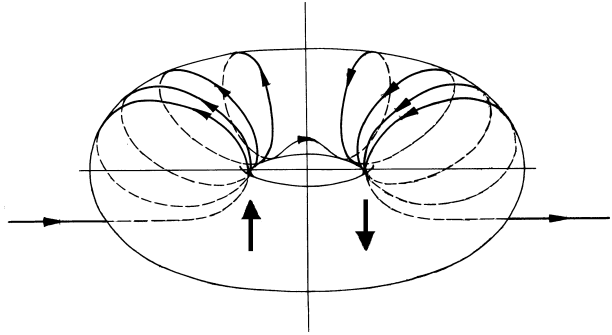}
\caption{Tverskoi's toroidal eddy winding a
magnetic field line. One toroidal surface of
the family forming the vortex is shown with a magnetic field line
that has accomplished four revolutions with fluid particles
moving over this surface. Two antiparallel heavy vertical arrows
mark the flux concentrations of opposite polarities
(slightly above).} \label{torvort}
\end{figure}

\section{Convective mechanism: basic idea and its implications}

To our knowledge, no local MHD dynamo mechanisms were
considered for 20 years since the study by \citet{gurleb}.
\citet{tve} was likely the first to suggest a convective
mechanism of local dynamo. He considered a simple kinematic
model describing the formation of a magnetic bipole by an
axisymmetric toroidal eddy in a perfectly conducting fluid,
which he considered a schematic representation of a convection
cell (more specifically, a supergranular cell was meant). The
action of Tverskoi's mechanism is illustrated in
Fig.~\ref{torvort}. The fluid particles move in circular
trajectories, concentric in any meridional section. Such a
velocity field is specified in an orthogonal coordinate system
$r,\ \varphi,\ \chi$, where $\varphi$ is azimuthal angle and
$r$ and $\chi$ are polar coordinates in the meridional plane
$\varphi=\const$. The fluid velocity is
$$V_r=V_\varphi=0,\quad V_\chi=V_0(r)[1+(r/a)\cos\chi]^{-1},$$
where $a$ is the radius of the circle $r=0$ and $V_0$ is an
appropriately chosen function such that $V_0(r)=0$ at any $r$
exceeding a certain value $r_0<a$. Obviously, the circles of a
given radius $r=\const$ form a toroidal surface. One such
surface is shown in Fig.~\ref{torvort}. The magnetic field
lines, which are initially straight and horizontal, are wound
by the fluid motion around the tori; one magnetic field line is
shown in the figure. The solution of the induction equation
obtained by Tverskoi contains a component that grows
monotonically with time. If $a-r_0\ll a$, the magnetic field
lines form two flux concentrations, with oppositely directed
magnetic fields, in the central part of the eddy. They are
marked in the figure with two heavy antiparallel vertical
arrows below them. The azimuthal arrangement of these
concentrations corresponds to the direction of the initial
field. Tverskoi's prediction was later qualitatively confirmed
by nonlinear numerical simulations of magnetoconvection in the
form of a pattern of hexagonal, B\'enard-type cells. In these
simulations, the flow was strongly stabilised by the conditions
of periodicity in the horizontal directions \citep{gconvmech,
dg}. The calculated magnetic field was strongly amplified and
formed a bipolar structure similar to that found by Tverskoi
develops in each convection cell. In addition, the vertical
component of the magnetic field was enhanced in the contact
zones between the hexagonal cells, i.e., in the intercellular
lanes.

In essence, Tverskoi has demonstrated that the topology of the
flow is very important in terms of the MHD effects of
convection. He conjectured that supergranular convection cells,
interacting with the azimuthal component of the global magnetic
field\footnote{In this section, since we discuss the properties
of \emph{toroidal} eddies, we do not use the commonly accepted
term \emph{toroidal magnetic field} to avoid terminological
confusions and prefer to speak about the global \emph{azimuthal
field}.} in nearly the same way as such eddies interact with
the horizontal initial field, could produce the magnetic fields
of bipolar sunspot groups. In this case, as can be seen from
Fig.~\ref{torvort}, the east--west orientation of the bipolar
magnetic structures would result from the east--west direction
of the global azimuthal field, with the same arrangement of
polarities as in the case of the rise of a tube. The action of
this mechanism would be controlled by the global azimuthal
magnetic field, which provides initial conditions for each
individual field-amplification event. Thus, Tverskoi's model
appears to be as successful as the rising-tube model in terms
of agreement with the global properties of solar activity such
as the Hale polarity law.

Other local dynamos, which are based on granular and
supergranular motions in the quiet photosphere, can account for
fairly disordered magnetic fields on very small scales
\citep{cattaneo,voeglschuessl}. In essence, only
\citet{steinnrdl} suggested a convective mechanism that can be
considered an alternative to the rising-tube model. They
numerically simulated the formation of an active region via the
flux rise due to convective motions. The computed scenario is
not related to the pre-existence of a coherent flux tube.
Initially, a uniform, untwisted, horizontal magnetic field is
present, which subsequently forms magnetic loops with a wide
range of scales. However, the initial field is required to be
fairly strong, and only a moderate magnetic-field
amplification, by a factor of about three, can be achieved.
Further studies of MHD convection in the context of local
dynamo were done, e.g., by \citet{kitiashetal}.

It is now clear, however, that Tverskoi's convective mechanism
in its original form can hardly describe the actual processes.
Without going into details, we only note the following: first,
the stability of the convection cell winding the magnetic field
lines is completely beyond the scope of the model, although
stability is of crucial importance for the efficiency of the
mechanism; second, well-developed sunspot groups are typically
larger than supergranules.

Nevertheless, Tverskoi's model, if it is modified in some way,
appears to be able to catch some important aspects of the
processes forming local magnetic fields. It can easily be
imagined that especially large and energetic cells sometimes
originate in the convection zone and, interacting with the
global azimuthal field, \emph{occasionally} give rise to strong
bipolar magnetic fields. It is in this respect that they
contrast with the ubiquitous ``normal'' supergranules, which
could be expected to produce smaller-scale fields. On the other
hand, the convective mechanism can also act on smaller spatial
scales, being responsible for the development of various local
magnetic features. It can also act in parallel with the process
of magnetic-field-line sweeping.

\begin{figure} 
\centering
\includegraphics[bb=0 0 400 360pt,clip,
width=0.45\textwidth]{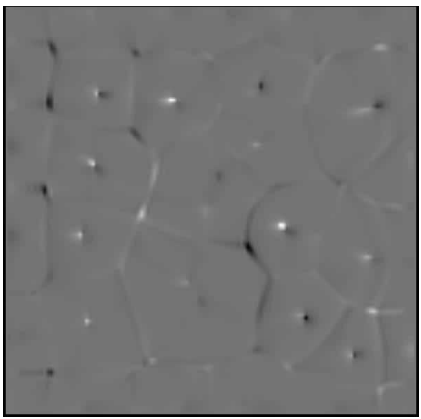}
\caption{Gray-scale map of the vertical magnetic-field component,
$B_z$, in a horizontal plane located near the no-slip upper boundary of the computation domain
in a simulation run for a Rayleigh number of $\mathrm{Ra}\approx 50\mathrm{Ra_c}$
(where $\mathrm{Ra_c}$ is the critical Rayleigh number), a Prandtl number of
$\mathrm{Pr}=30$, a magnetic Prandtl number of $\mathrm{Pr_m}=300$ and a Hartmann
number of $\mathrm{Ha}=0.01$ \citep{{gkm}}. The domain measures $8\times 8\times 1$.
The magnetic field is shown for a time of $10t_\nu$ ($t_\nu$ being the time
of viscous transfer across the layer); the magnetic-field strengths range from
$-28.5$ to 17.2 (in the units of the initial field strength); at the same time,
the field strength at the mid-height of the layer ranges from $-116.7$ to 162.1.}
\label{Hsimul}
\end{figure}

Some steps were made to modify Tverskoi's mechanism so as to
extend its area of applicability. Specifically, simulations of
magnetoconvection developing from random initial thermal
perturbations were carried out for a domain far exceeding the
expected characteristic convection wavelength in the horizontal
directions \citep[][see Fig.~\ref{Hsimul}]{gkm}. The initial
magnetic field was assumed to be uniform and horizontal. At the
initial evolutionary stage, a system of cells develops in the
form of irregular polygons. It was shown that cellular
magnetoconvection can produce bipolar (and also diverse more
complex) configurations of a substantially amplified magnetic
field. This occurs both in the inner parts of convection cells,
where magnetic field lines are ``wound'' by circulatory fluid
motion, and in the network formed by their peripheral regions
due to the ``sweeping'' of magnetic field lines. The topology
of the flow plays a fundamental role in the operation of this
mechanism, and it can be expected that the basic regularities
of the process should manifest themselves in nearly the same
way on different spatial scales.

Obviously, the manifestations of the rising-tube and convective
mechanisms should substantially differ. The point is to find
observational criteria to distinguish between the action of
either of them. With this aim in view, we developed an
observational program to study the evolution of both the
velocity and magnetic fields in growing active regions. This
program (operation plan) is intended for implementation with
the SOT on the \emph{Hinode} spacecraft
\citep{tsunetaetal,suematsuetal,shimizuetal} and has been
designated as HOP181
(\texttt{http://www.isas.jaxa.jp/home/solar/hinode\_op/%
hop.php?hop=0181}). It consists in simultaneous recording and
analysing the dynamics of the velocity and magnetic full-vector
fields on the photospheric level. We present here some
preliminary, purely qualitative results of processing the data
obtained at an initial implementation stage of the program.

\section{Observations and data processing}\label{obs}

The object of observations was the bipolar magnetic structure
that emerged within AR 11313 during the early evolution of this
structure, on 9--10 October 2011; the region was then near the
centre of the solar disc. Five 2-h-long observational sessions
were carried out with intervals that varied from 3 h 40 min to
6 h 30 min.

During each session, a $150''\times 163''$ field of view was
observed using the Narrowband Filter Imager (NFI) of the SOT at
two wavelength positions of FeI $\lambda$5776 \AA\ with a time
cadence of 2~min and a pixel size of $0.16''$. This yielded a
series of photospheric images, which can be used to calculate
horizontal velocities,\footnote{Since the area of interest was
located near the solar-disc centre and, moreover, corrections
for projection effects are not important from the standpoint of
our goal, we do not make difference here between the
transversal (tangential) and horizontal vector components and
also between the line-of-sight and the vertical component.} and
a series of Dopplergrams representing the line-of-sight
velocities. Simultaneously, the same FOV was scanned with the
Spectro-Polarimeter \citep[SP; see][]{ichimotoetal,litesetal13}
one or two times a session. The SP scan was done in the
so-called fast mode with a pixel size of $0.32''$, taking 32
min to obtain one SP map. To derive full-vector magnetic fields
from these SP observations, we used the MERLIN inversion code
\citep{litesetal07}, which assumed a Milne--Eddington
atmosphere. At this investigation stage, however, we did not
use data for the line-of-sight velocity and tangential magnetic
field.

The processing of the data included:
\begin{enumerate}
\item subsonic filtering based on Fast Fourier Transform;
\item constructing Dopplergrams;
\item an intensity-scaling procedure enhancing the image
    contrast by means of cutting off the tails of a
    pixel-intensity histogram and subsequent linear mapping
    of the remaining portion of the histogram to the whole
    admissible intensity range;
\item alignment of the images and Dopplergrams of each
    series with one another and, after a proper spatial
    scaling, with the magnetogram;
\item determination of the horizontal-velocity field using
    a technique based on the same principle as the standard
    local-correlation-tracking (LCT) method but more
    reliable \citep[see][for a description]{gbuch} and
    construction of cork-motion maps. This technique
    differs from the standard LCT procedure in a special
    choice of trial areas (``targets''), whose
    displacements are determined by maximizing the
    correlation between the original and various shifted
    positions of the target. Specifically, an area is
    chosen as a target in a certain neighborhood of each
    node of a predefined grid if either the contrast or the
    entropy of the brightness distribution reaches its
    maximum in this area. The horizontal velocities
    obtained are then interpolated to the positions of
    imaginary ``corks'' using the Delaunay triangulation
    and affine transformations specified by the deformation
    of the obtained triangles at the time step considered.
\end{enumerate}

The maps obtained at step (5) represent the trajectories of
imaginary ``corks'' that follow the velocity field inferred
from a series of images and attributed, to some approximation,
to the material flow. We compared the trajectory maps with the
maps of the magnetic field for times close to the mid-times of
the corresponding series.

\begin{figure} 
\centering
\includegraphics[bb=12 40 382 400pt,clip,width=0.45\textwidth]{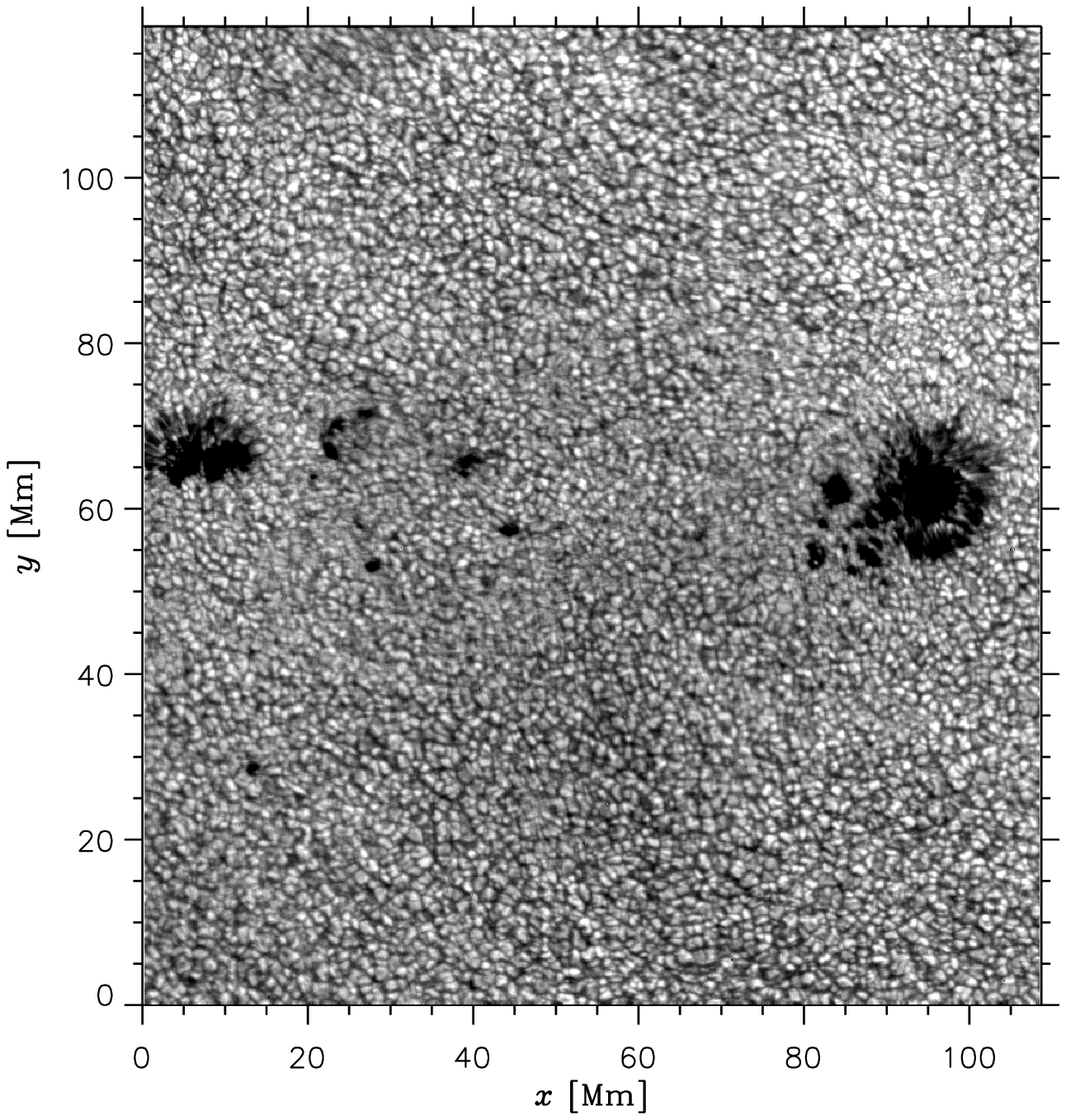}
\includegraphics[bb=12 40 382 400pt,clip,width=0.45\textwidth]{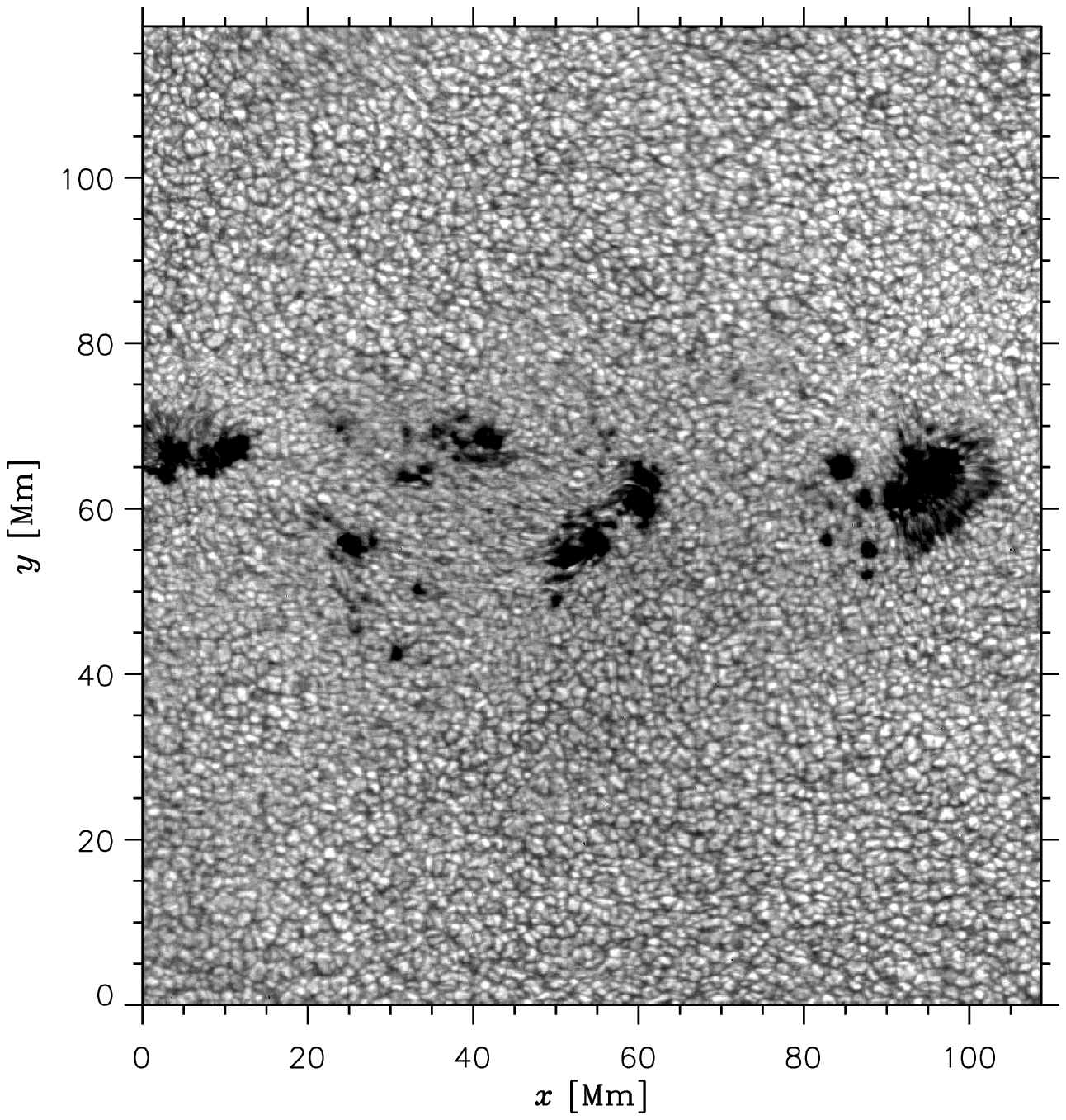}\\
\includegraphics[bb=12 40 382 400pt,clip,width=0.45\textwidth]{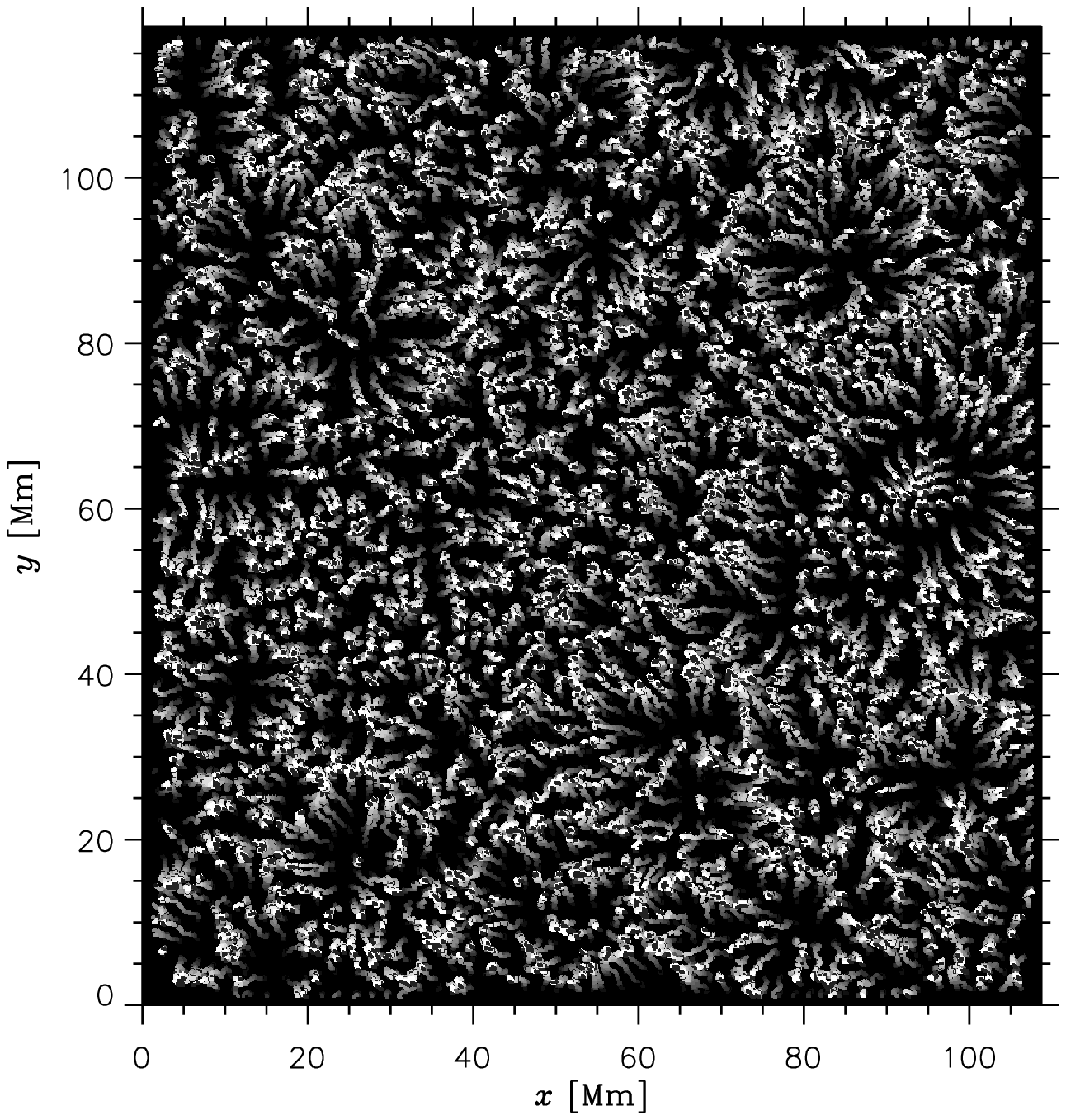}
\includegraphics[bb=12 40 382 400pt,clip,width=0.45\textwidth]{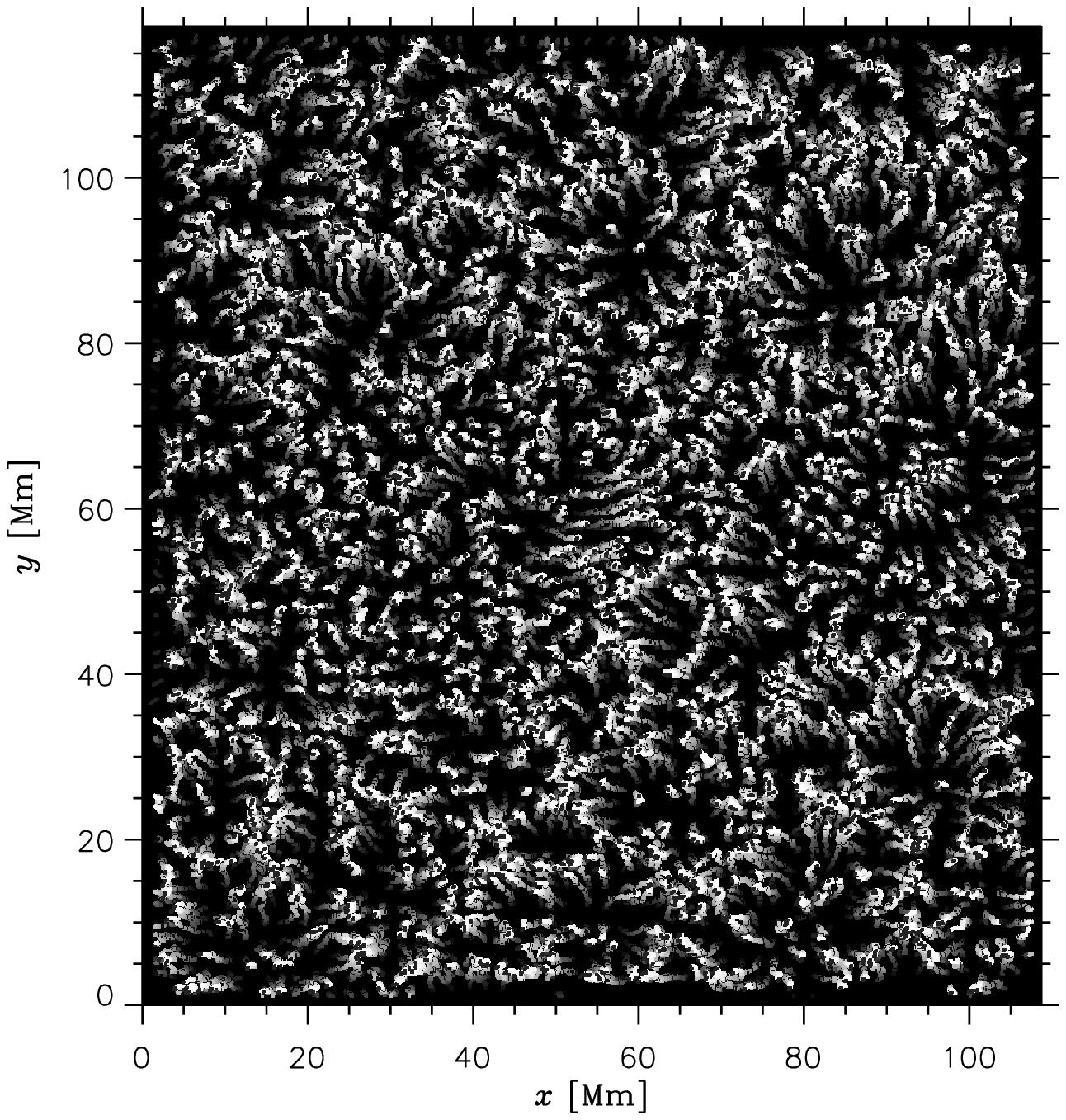}\\
\includegraphics[bb=12 12 382 400pt,clip,width=0.45\textwidth]{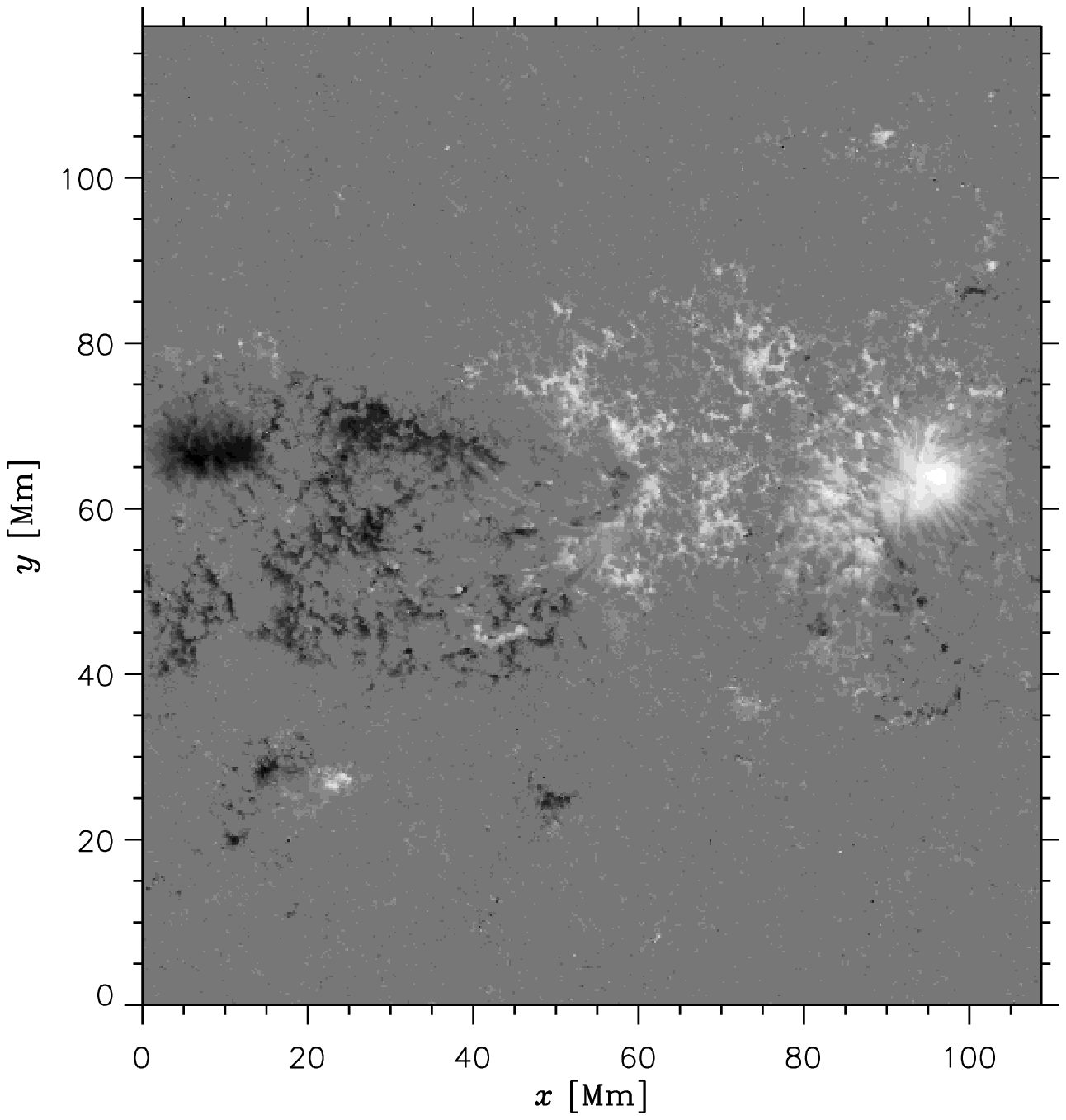}
\includegraphics[bb=12 12 382 400pt,clip,width=0.45\textwidth]{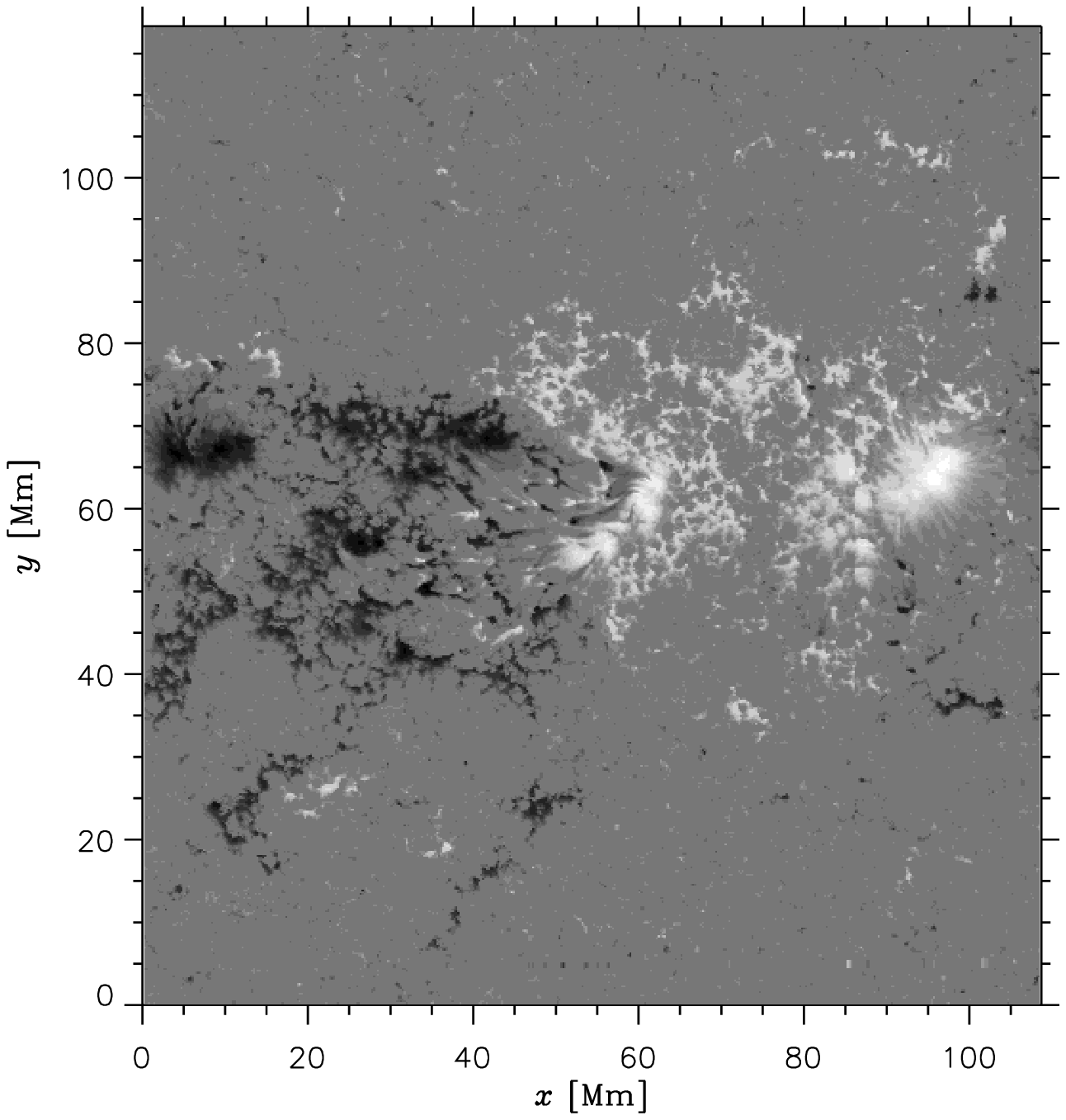}\\
\hspace{0.8cm}(a)\hspace{5.8cm}(b)
\caption{Comparison of intensity maps (top), horizontal-velocity fields (middle)
and line-of-sight magnetograms (bottom) obtained during the first (a) and third
(b) observational sessions (at 18--20$^\textrm h$ of 9 October and
06--08$^\textrm h$ of 10~October 2011, respectively; the intensity maps
and magnetograms were taken at the mid-times of these intervals). The trajectory of each cork
in the velocity map starts with a black dot and terminates at a bright white dot.
Light and dark areas in the magnetograms correspond to two signs of the magnetic field.}
\label{comp}
\end{figure}

Samples of the maps, which were qualitatively compared, are
shown in Fig.~\ref{comp}. For each session, a contrast-enhanced
FeI $\lambda$ 5576 \AA\ image, a cork-trajectory map and a map
of the line-of-sight magnetic field are presented. The
trajectory of each cork in a velocity map starts with a black
dot and terminates at a bright white dot. In most cases, the
2-h length of the session is sufficient for the corks to reach
stagnation segments of their trajectories, where the corks no
longer move over the photospheric surface. The stagnation
segments should obviously correspond to downflow areas in the
velocity field.

\section{Results}

As can be seen from a comparison between the top panels of
Figs~\ref{comp}a and \ref{comp}b, a fairly large bipolar
sunspot group has already formed in the area under study by the
time of the first observational session. At nearly the same
time, a new group starts developing between the main spots, in
the left half of the field of view. This process becomes mainly
accomplished by the third session (Fig.~\ref{comp}b).

\begin{figure} 
\centering
\includegraphics[bb=12 40 382 400pt,clip,width=0.75\textwidth]{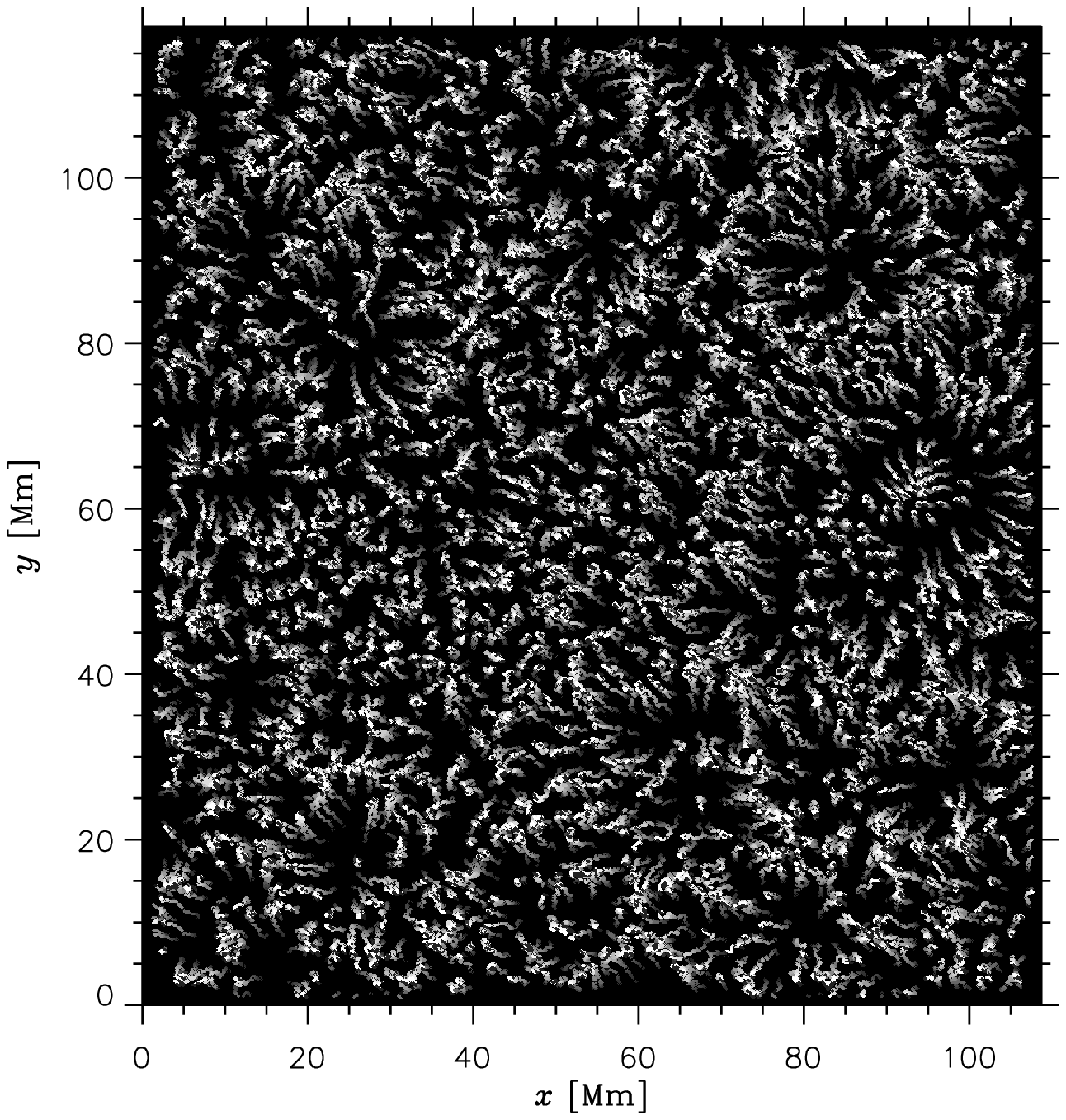}
\includegraphics[bb=12 40 382 400pt,clip,width=0.75\textwidth]
{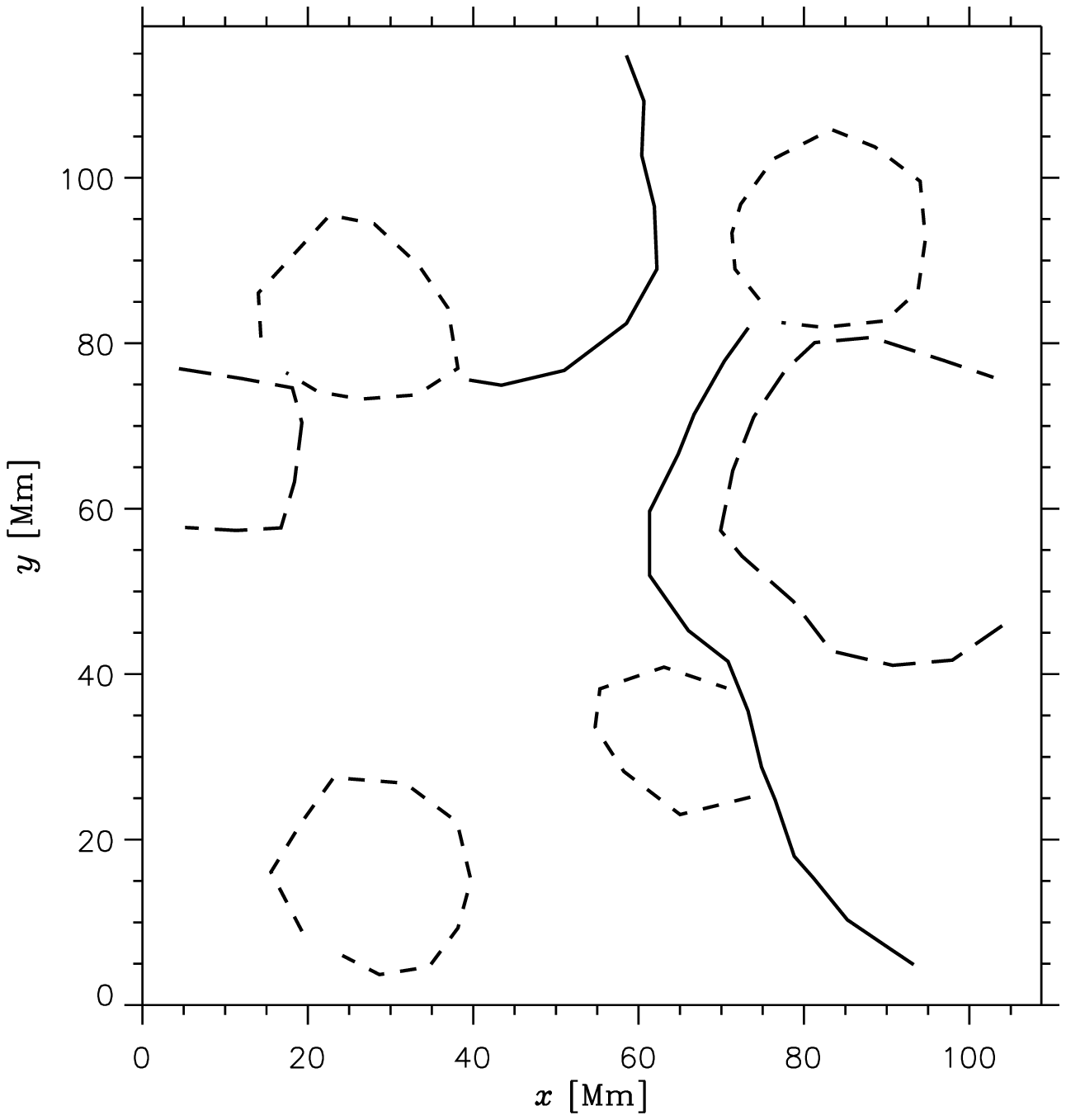}
\caption{Top: enlarged map of the horizontal-velocity field shown
in the middle panel of Fig.~\ref{comp}a; bottom: schematic outline of some features
present in this map, viz., lines of convergence (solid curves), contours of areas occupied by
sunspot-related divergent flows (long-dashed curves) and contours of most pronounced
supergranules (short-dashed curves).}
\label{enlarged}
\end{figure}

The middle panels of Figs~\ref{comp}a and \ref{comp}b represent
the velocity field by maps of cork trajectories. To make some
features of the flow better distinguishable, the velocity map
of Fig.~\ref{comp}a is additionally presented here in an
enlarged form in the top panel of Fig.~\ref{enlarged} together
with a schematic outline of the most pronounced features in the
bottom panel, both being plotted on the same scale. In the
velocity maps, several areas of local divergent flows
apparently corresponding to supergranules can be identified;
four of them are marked with closed short-dashed contours in
the bottom panel of Fig.~\ref{enlarged}. These
supergranule-sized flows are very similar in their appearance
to ``normal'' supergranular flows in the quiet photosphere as
detected using the same technique \citep[][Fig.~2]{gbuch}

A careful consideration of the two velocity maps reveals
neither any spreading flow on the scale of the developing group
nor any flows more intense than normal supergranular
convection. This is especially evident from the cork-trajectory
map of the first session (middle panel of Fig.~\ref{comp}a and
upper panel of Fig.~\ref{enlarged}), which corresponds to an
early development stage of the new sunspot group. At this
stage, a large-scale spreading flow would be well pronounced in
the case of the emergence of a strong flux tube. However, no
large-scale outflow from the location of developing group can
be found in the map; in contrast, local spreading flows on
meso- and supergranular scales unaffected by any large-scale
disturbances are observed.

As concerns the maps for the third session (Fig.~ \ref{comp}b),
a stream seemingly related to the newly formed magnetic feature
can be seen in the central part of the middle panel. It issues
from the vicinity of the point with coordinates of (50~Mm,
60~Mm) in the middle panel of Fig.~\ref{comp}b, is mainly
directed rightward and does not resemble a spreading flow. The
magnetogram in the bottom panel of the same figure exhibits
faint signs of small-scale cellular structuring of the magnetic
field on the left of the field-of-view centre, within the area
where the new active processes occur. Such structuring is
hardly compatible with the idea of flux-tube emergence. This is
additional evidence against the presence of a spreading flow
that could be associated with the emergence of an
$\Omega$-shaped loop of intense magnetic-flux tube, all the
more so because the emergence process would mainly have
terminated by that time.

In the velocity maps, bright curves stretching over long
distances can be distinguished, which appear as condensations
of the end segments of the cork trajectories approaching these
curves from two sides, i.e., as separatrices between oppositely
directed flows. Two such curves are shown as solid curves in
the scheme of Fig.~\ref{enlarged}, bottom. Quite likely, they
are merely due to aggregate visual effects of centrifugal flows
in different supergranules; however, even in this case, this
pattern disagrees with the velocity field that could be
expected in the case of emergence of an intense flux tube.

Apart from our discussion of the formation mechanisms for
strong local magnetic fields, it is interesting that our
trajectory maps reveal some details of flow structure related
to well-developed sunspots. Although the brightness
inhomogeneities in the umbras of the main spots are not
distinguishable by eye, they are nevertheless sufficient to
visualise the structure of the velocity field. As
Fig.~\ref{comp} indicates, there are flows converging to the
umbra centres. At the same time, diverging flows can be seen
around these spots.

\section{Summary and conclusion}

To summarise our qualitative findings, we note that the
velocity field in a growing active region has nothing to do
with a flow pattern that could be expected if an
intense-magnetic-flux tube emerged at the photospheric level.
This can be seen from (1) the absence of violent spreading
flows on the scale of the entire growing magnetic region, (2)
the preservation of normal mesogranules and supergranules, (3)
the signs of small-scale structuring of the magnetic field in
the area where new spots develop and (4) the signs of the
presence of separatrices between polarities, to which material
flows converge. In this context, it is interesting to remember
some inferences made very long ago.

As \citet{bumhow} wrote, the development of a spot group seems
to be controlled by the supergranular network: ``the ordered
motion within the supergranule is not only strong enough to
concentrate weak fields to the outer boundary, but also has
sufficient strength and scale to greatly influence the
positions of spots that have magnetic fields of at least
several hundred gauss. So the supergranular pattern is of
fundamental importance in the formation and structure of
individual active regions.'' In particular, the larger spots
occupy what appears to have been one or two supergranules.
Furthermore, \cite{bum63} has found that ``the photospheric
plasma moves approximately along the lines of force of the
intense local field.'' A comprehensive summary of these studies
was given by \citet{bum}, who noted that the development of an
active region does not involve the destruction of the
pre-existing convective-velocity field; in essence, the
observations demonstrate that the magnetic field coming from
below ``seeps'' through the network of convection cells.

In addition, we revert to the above-mentioned features of
magnetic-field evolution observed in the ``trilobite'' movie
\citep{nasa}. Diverging flows may be associated with individual
magnetic islands (the ``effervescent-tablet'' effect) rather
than with the whole region.

Our technique also reveals the flow structure in sunspots.
Specifically, we managed to detect converging streams in the
spot umbras, while spreading streams are observed around the
spots.

In conclusion, let us comparatively discuss the consistency of
the rising-tube and the convective model with observations. To
this end, we list here some points of disagreement between the
rising-tube model and observations, based on both the
observational data available in the literature and our
findings, and then demonstrate how the convective model avoids
these points.
\begin{itemize}
  \item The developing magnetic fields are observed to ``seep''
      through the photosphere without breaking down the existing
      supergranular and mesogranular velocity field, in contrast
      to what should be expected if a flux tube rose. In
      particular, a strong horizontal magnetic field at the apex
      of the rising loop should emerge on the surface and impart
      a roll-type structure to the convective flow.
  \item In contrast to the observed complex patterns of magnetic
      fields, this strong horizontal field would be a predominant
      magnetic feature on the scale of the entire active region
      before the origin of a sunspot group.
  \item No spreading flows are observed on the scale of the
      entire complex magnetic configuration of the
      developing sunspot group, as could be expected in the
      case of the emergence of a tube. Instead, flows are
      locally associated with each small-scale magnetic
      island. In particular, such finely structured flows
      (the ``effervescent-tablet'' effect) can be clearly
      seen in the ``trilobite'' magnetogram obtained on
      \emph{Hinode} \citep{nasa}.
  \item The presence of ``parasitic'' polarities within the
      area filled with a predominant magnetic polarity is
      not accounted for by the rising-tube model.
  \item The coexistence of differently directed vertical
      velocities inside the regions of a given magnetic
      polarity appears to be inconsistent with this model.
\end{itemize}
In our opinion, the convective mechanism appears to be in
better agreement with observational data in view of the
following:
\begin{itemize}
  \item The observed consistency of the developing magnetic field
      with the convective velocity field is an inherent property
      of this mechanism.
  \item Since the amplified magnetic field should largely
      be collinear with the streamlines, no strong
      horizontal field should connect different polarities.
  \item If convection forms local magnetic fields, spreading
      flows should actually be associated with developing
      magnetic islands rather than with the entire complex.
  \item Diverse complex patterns with mixed polarities can
      be accounted for in a natural way by the presence of
      a fine structure of the convective flow.
  \item The convective mechanism can in principle operate on
      various spatial scales, being controlled solely by the
      topology of the flow.
\end{itemize}

Strictly speaking, our inferences from the described
observational data may not be quite universal, since we have
studied only one evolving sunspot group, which, in addition,
developed within an already formed group. Therefore, care must
be taken in extending our findings to the overall pattern of
MHD processes. Acquisition of more observational information in
parallel with further development of data-processing techniques
is needed to make more reliable conclusions concerning the
formation mechanisms for the photospheric magnetic fields.

Nevertheless, the observational data described here can hardly be
interpreted in the framework of the rising-tube model, so that the
convective model appears to be more promising in terms of the
representation of reality.

\section*{Acknowledgments}

\emph{Hinode} is a Japanese mission developed and launched by
ISAS/JAXA, with NAOJ as domestic partner and NASA and STFC (UK)
as international partners. It is operated by these agencies in
cooperation with ESA and NSC (Norway).

\end{document}